\documentclass[sigconf,natbib=false]{acmart}

\AtBeginDocument{%
  }

\setcopyright{acmlicensed}
\copyrightyear{2025}
\acmYear{2025}
\acmPrice{}
\acmISBN{}
\acmDOI{}

\acmConference[SIGSPATIAL ’25]{The 33nd ACM International Conference on Advances in Geographic Information Systems}{November 03--06, 2025}{Minneapolis, MN}
\RequirePackage[
  datamodel=acmdatamodel,
  style=acmnumeric,
  ]{biblatex}

\usepackage{array}    
\usepackage{booktabs} 
\usepackage{multirow} 
\usepackage{pgfplots}
\pgfplotsset{compat=1.17}
\usepackage{xcolor}
\usepackage{wrapfig}

\definecolor{bblue}{HTML}{4F81BD}
\definecolor{rred}{HTML}{C0504D}
\definecolor{ggreen}{HTML}{9BBB59}
\definecolor{ppurple}{HTML}{9F4C7C}

\pgfplotsset{
	/pgfplots/my legend/.style={
		legend image code/.code={
			\draw[thick,black](-0.05cm,0cm) -- (0.3cm,0cm);%
		}
	}
}

\begin{document}

\title{MapLibre Tile: A Next Generation Vector Tile Format}

\author{Markus Tremmel}
\email{markus.tremmel@rohde-schwarz.com}
\orcid{}
\affiliation{%
  \institution{Rohde \& Schwarz GmbH \& Co. KG}
  \city{Teisnach}
  \state{Bavaria}
  \country{Germany}
}

\author{Roland Zink}
\email{roland.zink@th-deg.de}
\affiliation{%
  \institution{Deggendorf Institute of Technology}
  \city{Deggendorf}
  \state{Bavaria}
  \country{Germany}
}

\renewcommand{\shortauthors}{Tremmel et al.}

\begin{abstract}
	
The Mapbox Vector Tile (MVT) format is widely considered the leading open standard for large-scale map visualization, as evidenced by its widespread adoption by major technology companies such as AWS, Meta, and Microsoft for their products and services. However, MVT was developed nearly a decade ago and, consequently, does not fully align with the capabilities of new geospatial data sources that are characterized by rapidly increasing data volumes due to advancements in geospatial sensors and automated detection through artificial intelligence. In this paper, we introduce the MapLibre Tile (MLT) format, a novel vector tile specification designed from the ground up to address the limitations of MVT. Our experiments, simulating user sessions on widely used basemap datasets, demonstrate that MLT achieves up to three times better compression ratios compared to MVT on encoded tilesets, with over six times better on certain large tiles. Additionally, MLT offers decoding speeds that are up to three times faster and significantly enhances processing performance. MLT also introduces new functionalities and is specifically designed to lay the foundation for the next generation of map renderers, which we expect to entirely offload processing to the GPU, thereby overcoming the stagnation of Moore’s law.

\end{abstract}





\maketitle

%TODO: remove again when ACM Reference Format is valid
%\newpage

\section{Introduction}

The Mapbox Vector Tile (MVT) format is widely recognized as the de facto standard for low-latency visualization of large-scale maps \cite{fujimura_design_2019}. Nevertheless, its design no longer fully aligns with the capabilities of contemporary hardware, modern APIs, and the evolving requirements of emerging geospatial data sources which are characterized by rapidly growing data volumes due to automated detection powered by artificial intelligence. To address these challenges, we gathered a comprehensive set of functional and non-functional requirements for a new open vector tile format, collaborating with various stakeholders from the community, primarily through the MapLibre organization, as well as the industry, including AWS, Microsoft, and Rohde \& Schwarz. Additionally, we conducted an analysis of the latest emerging geospatial data sources, including those from Overture Maps, and reviewed engineering reports on the Vector Tiles Pilot from the Open Geospatial Consortium (OGC). 

This paper introduces the MapLibre Tile (MLT) format, a novel vector tile specification specifically designed to enable low-latency rendering of massive geospatial datasets. In comparison to MVT, MLT significantly reduces overall tile sizes while simultaneously improving decoding and processing performance. Furthermore, MLT enhances functionality by supporting a complex type system, incorporating nested types such as lists and structs, managing 3D coordinates, and integrating m-coordinates and linear referencing.

\section{MapLibre Tile Format}

\textbf{Conceptual Model.} To abstract real-world phenomena such as roads or buildings, MLT utilizes the well-established concept of a feature data structure. Each feature is defined by a unique identifier (ID), a geometric property that describes its position and shape, and a set of attributes representing the associated non-spatial data. The geometry model of MLT is conceptually based on the Simple Feature Access (SFA) standard \cite{consortium_simple_2017} of the OGC, with the exception of support for the GeometryCollection type. To efficiently encode multiple per-vertex attributes, known as m-coordinates, along with per-feature attributes that provide a single value for an entire geometry, the attributes are classified into feature-scoped and vertex-scoped. A set of features with the same thematic reference that share a common set of attributes and typically the same geometry type is organized into a FeatureTable. Based on their geometric properties, features are subdivided into a grid of smaller, spatially uniform partitions known as tiles. Since MLT is optimized for rendering workloads, the source geometries of a feature within a tile are quantized into integer screen coordinates defined by a grid coordinate system, facilitating more space-efficient encoding, adapted from MVT.

\textbf{Physical Model.} To achieve efficient storage and low-latency network transfer while ensuring rapid processing performance at runtime, MLT is structured into both a storage format and an in-memory format. This dual approach is inspired by leading big data analytic formats, specifically Apache Parquet \cite{vohra_apache_2016} and Apache Arrow \cite{the_apache_software_foundation_apache_2025}. By tightly coupling the storage and in-memory representations, MLT achieves rapid decoding speeds, in line with recommendations of \cite{zeng_empirical_2023}. Both formats of MLT employ a column-oriented storage model, also referred to as the Decomposition Storage Model (DSM), in contrast to the record-oriented approach utilized by MVT. A column-oriented layout offers the advantages of column-specific compression, vectorized execution, improved cache utilization, and enhanced CPU efficiency compared to a row-oriented model \cite{abadi_design_2013}. In the context of vector tiles, the drawbacks associated with a columnar layout, such as limited transaction support and higher record assembly costs, are largely negligible. This is primarily due to the fact that tiles are typically fully regenerated, and modern map renderers rarely require the complete reconstruction of features.

\subsection{Storage Format}

\textbf{Structural Encoding.} For the structural encoding of nullable and variable-size geometry and attribute types of features into a columnar layout, a length/presence pair-based encoding \cite{zeng_empirical_2023} is employed. Consequently, a logical column is divided into multiple physical subcolumns, referred to as streams, that are stored adjacently. A stream consists of a sequence of values in a continuous memory block, all sharing the same type, accompanied by metadata such as size and encoding type. To achieve space-efficient encoding, streams are categorized into four types: present, offset, length, and data streams. For instance, a dictionary-encoded nullable string attribute column may include a present stream indicating value presence, an offset stream referencing the position within the data stream where a particular value can be found, a length stream detailing the character count of each string, and a data stream containing the actual UTF-8 encoded string values.

%TODO: use ordered set notation -> ⟨(𝑥1, 𝑦1), (𝑥2, 𝑦2), . . ., (𝑥𝑛, 𝑦𝑛)⟩ 
\textbf{Geometry Encoding.} At its core, MLT stores all coordinates of the geometries interleaved within a single contiguous buffer per FeatureTable, referred to as the VertexBuffer, supporting both 2D  \(\{(x_1, y_1), (x_2, y_2), \ldots, (x_n, y_n)\}\) and 3D coordinates \(\{(x_1, y_1, z_1), \\ (x_2, y_2, z_2), \ldots, (x_n, y_n, z_n)\}\). To define the connectivity between the vertices and to support the geometry model of the SFA standard, a hierarchical data structure with up to three levels of nesting is employed, depending on the geometry type. Consequently, the aforementioned structural encoding is implemented across three distinct topology streams: (i) the Geometries stream, which indicates the number of geometries per feature for multipart geometries; (ii) the Rings stream, which specifies the number of LinearRings when encoding polygon types; and (iii) the Vertices stream, which encodes the number of vertices for each LineString or LinearRing. To further support mixed geometry types within a single geometry column, MLT incorporates a Type stream that specifies the geometry type of each feature, enabling efficient encoding of various geometries as a union type.

A notable feature of MLT is the storage of pre-tessellated polygon meshes directly within the file. This approach allows the computationally intensive triangulation step during runtime (online tessellation), often considered a major bottleneck in modern GPU-based map rendering, to be offloaded to the tile generation phase (offline tessellation). This is accomplished by utilizing an additional IndexBuffer stream that stores the triangle indices of polygons, along with an optional Triangles stream when a complete reconstruction of the features is necessary. As a result, these geometries can be transferred directly to the GPU, often without any further processing, thereby significantly enhancing the runtime performance of vector tile rendering.

\begin{figure*}[h]
	\centering
	\includegraphics[width=\linewidth]{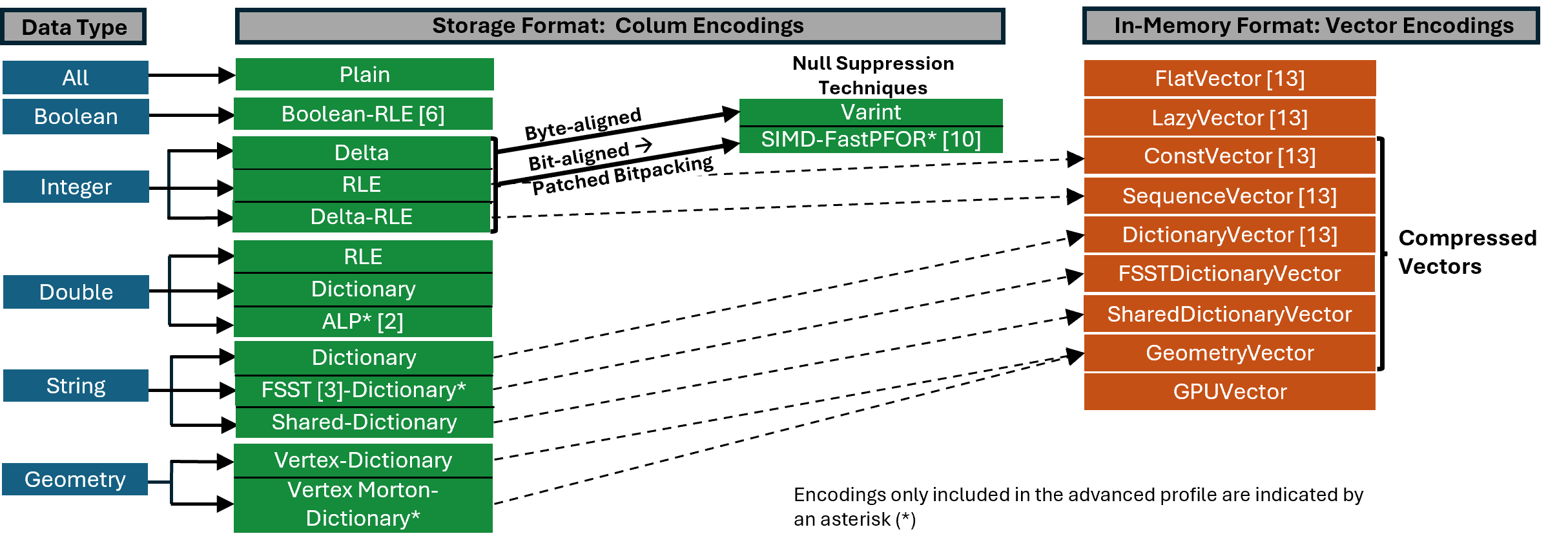}
	\caption{Encoding pool for the storage and in-memory format \cite{the_apache_software_foundation_orc_2025, lemire_decoding_2015, afroozeh_alp_2023, boncz_fsst_2020, pedreira_velox_2022}}
	\Description{Encodings.}
	\label{fig:encodings}
\end{figure*}

\textbf{Encoding Schemes.} 
To minimize the overall file size while enabling rapid decoding, MLT employs various type-specific lightweight encoding schemes, particularly optimized for efficient use in web browsers. The encoding pool, illustrated in the center of Figure \ref{fig:encodings}, was selected based on a comprehensive experimental evaluation by applying the encodings to various basemap datasets, primarily OpenStreetMap and Overture Maps, and analyzing the resulting compression ratio alongside with the decoding performance. The evaluation results further demonstrate the effectiveness on the compression ratio by recursively cascading encodings under specific criteria, a method known as hybrid encoding, as exemplified by formats like BtrBlocks \cite{kuschewski_btrblocks_2023}. In contrast to BtrBlocks, MLT permits only a fixed combination of encodings and restricts certain encodings to specific data types, to simplify the decoder implementation and ensure rapid decoding performance, as recommended by \cite{zeng_empirical_2023}. Additionally, to facilitate broader implementation of MLT, the encoding pool is divided into a simple profile and an advanced profile.

To achieve efficient coordinate compression, we introduce a novel geometry encoding algorithm that employs dictionary encoding for vertex coordinates, followed by dimension reduction through the ordering of these coordinates along a Z-order curve. To further compress the resulting Morton codes, we apply delta encoding in conjunction with the SIMD-FastPFOR algorithm \cite{lemire_decoding_2015} for effective null suppression. This encoding is particularly well-suited for use on the GPU, as it minimizes data transfer size from the CPU to the GPU, and the decoding overhead for the Morton codes becomes negligible due to the highly parallel processing capabilities of modern GPUs. 

\subsection{In-Memory Format}

The primary objective of introducing an explicit in-memory format is to enable rapid decoding from the storage format while optimizing processing performance, particularly for memory-bound map rendering tasks characterized by a low computation-to-data ratio, such as filtering operations and simple geometric computations. To achieve this, the format is designed to efficiently utilize underlying CPU caches and enable constant-time random access to arbitrary data. This is accomplished by transforming the columns of the storage format into vectors, as illustrated in Figure \ref{fig:encodings}, a concept inspired by Velox \cite{pedreira_velox_2022}. The transformation follows a three-step process: (i) decoding columns when opaque encodings are applied, (ii) converting the compact layout of nullable columns to a placeholder layout by filling in null values, and (iii) transforming the length information of variable-size types, such as strings, to offset values. Importantly, columns compressed with transparent encodings \cite{pace_lance_2025}, such as Dictionary or Delta-RLE, can be seamlessly converted into compressed vectors. By directly operating on this compressed data, we eliminate the decoding overhead and enhance cache locality, thereby facilitating faster processing and more efficient transfers to the GPU. To perform filtering operations on these vectors, we propose a vectorized query execution model in conjunction with a selection vector to identify valid tuples.

Furthermore, to enhance the performance of state-of-the-art map renderers on CPU-bound tasks such as line tessellation and font shaping, the in-memory format is designed for direct compatibility with modern parallel computing technologies, including SIMD and GPU compute shaders, without requiring additional data transformations. Notably, the efficient support of the novel WebGPU compute shader technology is expected to enable next-generation map renderers to adopt GPU-driven rendering approaches within the browser, effectively addressing the challenges associated with the stagnation of Moore's Law \cite{leiserson_theres_2020}. The seamless transfer to GPU buffers is enabled by three principal design features: (i) vectors are organized contiguously in memory using multiple fixed-size buffers, such as ArrayBuffer data structures in the Web; (ii) random access to arbitrary data is facilitated through an offset-based approach for nested types, rather than a pointer-based method; and (iii) data are aligned by powers of 2, in accordance with modern graphics APIs, such as ensuring that 3D integer coordinates are aligned to 16 bytes.

\section{Experiments}

This section presents the results of an experimental evaluation comparing the MLT and MVT formats using real-world datasets across three distinct categories: (i) the size of the formats, both encoded and compressed; (ii) the performance of decoding the encoded formats into their respective in-memory representations; and (iii) the performance of applying filtering operations on these in-memory representations. 
  
\textbf{Data Sets.} For this evaluation, a selection of widely used basemap datasets, available in the MVT format and exhibiting diverse characteristics, was chosen. Two tileset variants of an OpenStreetMap dataset, based on the widely adopted OpenMapTiles schema, were generated using different versions of Planetiler \footnote{https://github.com/onthegomap/planetiler.}. The first variant (OT1) was created with an earlier version 0.6.0 of Planetiler, representing a dataset without schema-specific optimizations for vector tiles. In contrast, the second variant (OT2) was generated with a newer version 0.8.3, representing a highly optimized tileset with extensively crafted layer-specific optimizations for that particular vector tile schema, aimed at mitigating the limitations of MVT. Additionally, we selected a tileset from Swisstopo (ST) \footnote{https://www.swisstopo.admin.ch/en/web-maps-base-map.}, which serves as a geospatial base dataset provided by a national surveying authority, as well as a dataset from Overture Maps (OV) \footnote{https://explore.overturemaps.org.}, representing emerging geospatial source formats that integrate AI-captured data in combination with nested encoded feature properties.  

\textbf{Methodology.} For benchmarking, tiles were selected by simulating user interactions with map applications, specifically by zooming from level 0 to 18 into the center of a major European city. This approach is inspired by the methodology outlined in \cite{netek_performance_2020} and has been adapted to ensure that all zoom levels contribute evenly to the benchmarking results. For OT1 and OT2, we simulate zooming into Munich; for Overture Maps, into London; and for Swisstopo, into Zurich. All experiments were executed on a machine equipped with a 1.9 GHz Intel Core i7-8665U processor, 32 GB RAM, and a 1 TB NVMe SSD disk drive, running Windows 11. To ensure the reproducibility of the findings, both the code and data for the following experiments are publicly accessible \footnote{https://github.com/mactrem/mlt-evaluation.}.

\subsection{Encoding \& Compression Performance}

As shown in Table \ref{tab:tile_size}, MLT achieves a significant reduction in size for encoded tiles compared to MVT across all datasets, with a maximum reduction by a factor over six for some large tiles. Specifically, for OT1 and ST, the size of MLT is about one-third that of MVT, while for OV, it is less than half. Interestingly, when applying an additional general-purpose compression algorithm (Gzip) on top of the lightweight encoded tiles, the simple encoding profile exhibits enhanced compression ratios. Notably, in all cases, the lightweight encoded MLT files are smaller than the additionally Gzip-compressed MVT files. Given that MLT already achieves significant size reduction with its lightweight encodings, the additional relative gains from applying heavyweight compression algorithms \cite{kuschewski_btrblocks_2023} are not as substantial as compared to MVT. This finding suggests that applying further heavyweight compression on MLT may be unnecessary in environments with fast network speeds, as additional size reduction gains are often negligible. By eliminating the need for heavyweight compression, significant improvements in decoding performance and energy efficiency can be achieved, aligning with trends and recommendations in data analytics \cite{zeng_empirical_2023, kuschewski_btrblocks_2023}. 

\begin{table}[htbp]
	\caption{Encoding and compression benchmarks (in MB), including maximum per-tile size reduction ratios (Max SR), with advanced profiles listed before simple profiles}
	\label{tab:tile_size}
	\centering
	\begin{tabular}{|p{1.13cm}|p{1.38cm}|p{1.38cm}|p{1.38cm}|p{1.38cm}|}
		\hline
		\textbf{Dataset} & \textbf{OT1} & \textbf{OT2} & \textbf{ST} & \textbf{OV} \\
		\hline
		& \multicolumn{4}{l|}{\textbf{Encoded}} \\	
		\hline
		MLT & 15.7 | 18.1 & 13.8 | 16.6 & 8.0 | 9.0 & 97.5 | 127.7 \\
		\hline
		MVT & 45.5 & 21.7 & 24.7 & 241.1 \\
		\hline
		SR & \textbf{2.9} | 2.5 & \textbf{1.6} | 1.3 & \textbf{3.1} | 2.8 & \textbf{2.5} | 1.9 \\
		\hline
		Max SR & 5.0 & 2.4 & 6.7 & 4.8   \\
		\hline
	\end{tabular}
	\vspace{0.05cm}
	\begin{tabular}{|p{1.13cm}|p{1.38cm}|p{1.38cm}|p{1.38cm}|p{1.38cm}|}
		\hline
		& \multicolumn{4}{l|}{\textbf{Compressed}} \\ 
		\hline
		MLT & 13.0 | 10.9 & 12.8 | 11.8 & 7.2 | 6.7 & 80.8 | 78.9 \\
		\hline
		MVT & 21.3 & 14.3 & 9.6 & 103.6 \\
		\hline
		SR & 1.65 | \textbf{1.96} & 1.12 | \textbf{1.22} & 1.35 | \textbf{1.44} & 1.28 | \textbf{1.31} \\
		\hline
	\end{tabular}
\end{table}

\subsection{Decoding \& Filtering Performance}

Given that the performance in browser-based environments is a key requirement, the following benchmarks are performed in JavaScript. First, we evaluate the performance of decoding from the storage format into the corresponding in-memory representation for MVT by using the vector-tile-js library \footnote{https://github.com/mapbox/vector-tile-js.}, version 2.0.3, and for MLT utilizing the related TypeScript decoder. Due to current limitations in the MLT TypeScript decoder, which does not yet support all advanced encodings, benchmarks are conducted on MLT files encoded with the simple encoding profile. As shown in Table \ref{tab:performance}, MLT consistently outperforms MVT by a factor of 2 to over 3, depending on dataset complexity, even without employing SIMD instructions. 

Finally, we evaluate the performance of applying filtering operations on the in-memory representations, a prevalent task in the processing of vector tiles to determine which features are rendered in a style layer. The filter expressions are based on the popular OpenMapTiles basic style \footnote{https://github.com/openmaptiles/maptiler-basic-gl-style.}, with slight modifications in the number of filters (35 out of 47) to ensure a fair comparison. We employ the FeatureFilter API from the maplibre-gl-style-spec library \footnote{https://github.com/maplibre/maplibre-style-spec.} for MVT that is based on a tuple(feature)-at-a-time query execution model (Volcano iterator model) and a modified version of the vectorized query execution model for MLT. As illustrated in Table \ref{tab:performance}, MLT outperforms MVT by a factor ranging from 3.7 to 4.4 in the real-world filtering use case.

\begin{table}[htbp]
	\caption{Decoding and filtering performance with the time measured in milliseconds.}
	\label{tab:performance}
	\centering
	\begin{tabular}{|p{1.13cm}|p{0.78cm}|p{0.78cm}|p{0.78cm}|p{0.78cm}| |p{0.78cm}|p{0.78cm}|}
		\hline
		& \multicolumn{4}{c||}{\textbf{Decoding}} &  \multicolumn{2}{c|}{\textbf{Filtering}} \\
		\hline
		Dataset & \textbf{OT1} & \textbf{OT2} & \textbf{ST} & \textbf{OV} & \textbf{OT1} & \textbf{OT2} \\
		\hline
		MLT & 397  & 280  & 258 & 1248  & 661 & 32 \\
		\hline
		MVT & 1212 & 554 & 636 & 3665 & 2845 & 117 \\
		\hline
		Speed up & 3.1  & 2.0  & 2.5  & 2.9  & 4.4 & 3.7 \\
		\hline
	\end{tabular}
\end{table}

\section{Conclusion and Future Work}

This paper introduces the MapLibre Tile format, a novel vector tile specification that is highly optimized for low-latency map rendering. Our benchmarks, conducted using four real-world datasets, demonstrate that MLT achieves a Pareto improvement in both compression ratio and decoding speed compared to the MVT format. These improvements are particularly significant for large tilesets, eliminating the need for complex vector tile schema-specific optimizations. 

In addition to its ongoing integration into the state-of-the-art open-source map renderers MapLibre GL JS and MapLibre Native, we are exploring GPU-driven rendering approaches for vector tiles to further enhance performance and fully leverage the capabilities of MLT. This initiative is part of our broader research objective to redesign the entire map rendering stack \cite{tremmel_markus_design_2025}, from efficient cloud-based map tiles deployment to high-performance GPU rendering. 

\printbibliography

\end{document}